\documentclass[10pt,  letter, twocolumn]{article}

\usepackage{authblk}

\usepackage{graphicx}
\usepackage[margin=0.75 in]{geometry}
\usepackage{hyperref}
\usepackage{color}

\usepackage{caption}
\title{High Impedance Detector Arrays for Magnetic Resonance}
\date{}

\author[1,2]{Bei Zhang\thanks{Bei.Zhang@nyumc.org}}
\author[1,2,3]{Daniel K. Sodickson}
\author[1,2]{Martijn A. Cloos}
 
\affil[1]{\tiny Bernard and Irene Schwartz Center for Biomedical Imaging, New York University School of Medicine, New York, NY, USA.}

\affil[2]{\tiny Center for Advanced Imaging Innovation and Research (CAI$^2$R), New York University School of Medicine, New York, NY, USA.}

\affil[3]{\tiny The Sackler Institute of Graduate Biomedical Sciences, New York University School of Medicine, New York, New York, USA.}

\begin{document}
    \maketitle
    
\textbf{Resonant inductive coupling is commonly seen as an undesired fundamental phenomenon emergent in densely packed resonant structures, such as nuclear magnetic resonance phased array detectors. The need to mitigate coupling imposes rigid constraints on the detector design, impeding performance and limiting the scope of magnetic resonance experiments. Here we introduce a high impedance detector design, which can cloak itself from electrodynamic interactions with neighboring elements. We verify experimentally that the high impedance detectors do not suffer from signal-to-noise degradation mechanisms observed with traditional low impedance elements. Using this new-found robustness, we demonstrate an adaptive wearable detector array for magnetic resonance imaging of the hand. The unique properties of the detector glove reveal new pathways to study the biomechanics of soft tissues, and exemplify the enabling potential of high-impedance detectors for a wide range of demanding applications that are not well suited to traditional coil designs.}
\vspace{0.5cm}   
  
Founded on the principles of nuclear magnetic resonance (NMR)\cite{Bloch1948}, magnetic resonance imaging (MRI)\cite{Lauterbur1973,Mansfield1973} is a pre-eminent clinical imaging modality and invaluable research tool. The rich information sought after in the NMR experiment is locked away deep in the interaction of the nuclei with one another and with the applied external magnetic field. The key to extracting this information is the deliberate perturbation of nuclear spins, and careful tracking of their response -- a sequential process reliant on radiofrequency (RF) magnetic fields, generated and observed by RF coils tuned to the Larmor frequency of the nuclear spins in question.

Modern-day magnetic resonance (MR) systems use phased array coils constructed out of low impedance resonant loops\cite{Roemer1990}. Inside such arrays, electrodynamic interactions between elements must be carefully balanced. Neighboring coil elements are strategically overlapped to minimize inductive coupling, and preamplifier interfaces are tuned to suppress the induced current on each element (Fig. 1A). This electromechanical balancing act becomes increasingly difficult as the number of receive elements grows to approach optimal performance\cite{Wang1995,Ocali1998,Schnell2000,Ohliger2003,Wiesinger2004,Lattanzi2010,Lattanzi2012}, leading to geometrical puzzles of profound complexity\cite{ Wiggins2009,Schmitt2008,Fujita2013}. Any change in coil structure disturbs the delicate balance, thus demanding rigid structures which retain the critical overlap, and which, therefore, cannot adapt to the variable needs of individual subjects, or provide the flexibility needed to study moving joints or other body kinematics.

\begin{figure}[h!]
\includegraphics[width=8.3cm ]{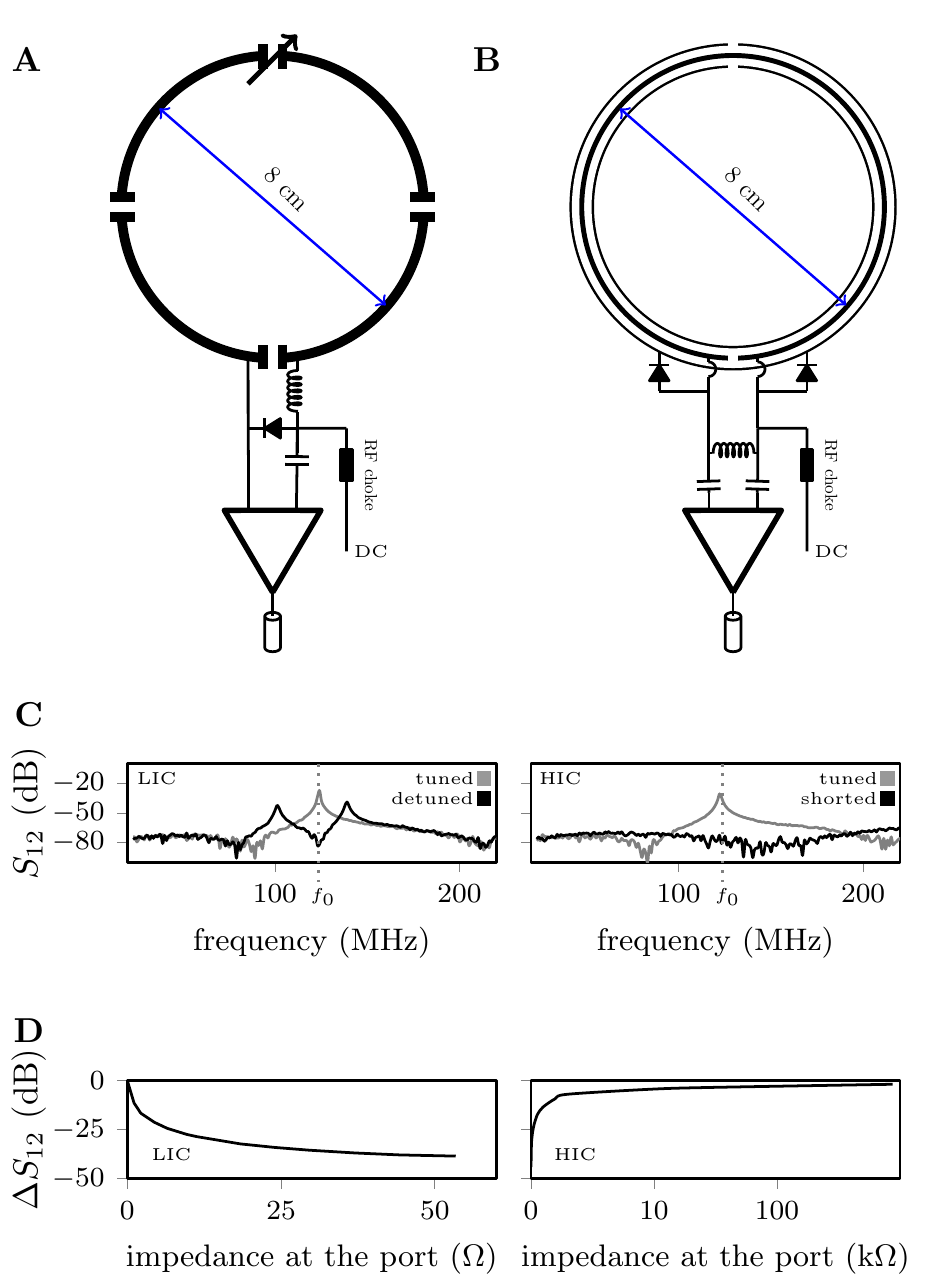}
\caption{Schematic drawing of a traditional LIC (A) and HIC (B) element. The resonances observed with a double probe experiment when the coil is tuned and detuned (C). The relative amplitude ($\Delta S_{12}$) of the induced current measured at the resonance frequency ($f_0$) during a double probe experiment as a function of the impedance at the port (D).
}\end{figure}

Here we show that a high impedance coil (HIC) design can eliminate all electrodynamic interactions between receive elements in a phased array coil. First, we introduce the HIC design concept and characterize its properties. We then present an exemplary application, the wearable ``glove coil'', which conforms to the shape of the hand, and enables the study of soft tissue mechanics as the hand moves freely inside the MR system, allowing us to see how the intricate arrangement of muscles and ligaments in the hand changes during complex tasks.

The key function of inductive detectors is to measure the electromotive force (EMF) induced by the NMR signal. Traditional low impedance coils (LIC) efficiently capture the EMF, but also allow current to flow. This current, in turn, creates a secondary RF field that can be detected by neighboring elements. To eliminate resonant inductive coupling between elements we set out to develop a new element design, which allows the EMF to be measured without allowing current to flow and signal to leak inductively into neighboring elements.

Inspired by developments in wireless power transfer systems \cite{Kurs2007,Tierney2014}, we designed a high impedance resonant coaxial NMR probe (Fig. 1B). In contrast to conventional LIC elements, there are no lumped capacitors to adjust the resonance frequency or distribute the current. Instead, our HICs were tuned by adjusting the length, the relative permittivity of the substrate, and the ratio between inner and outer conductor diameter\cite{selfMethods}, such that the Larmor frequency of interest coincides with the lowest resonance of the structure. In this situation, EM simulations (CST microwave studio, Darmstadt, Germany) reveal that the current on the inner conductor increases approximately linearly from zero at the gap to a maximum at the gap of the outer conductor, which is mirrored by an opposing current distribution on the inner surface of the outer conductor. At the gap in the outer conductor, skin depth effects allow a current to flow through the gap onto the outer surface of the conductive wrapping of the coil, and this current travels uniformly towards the other edge of the gap (Fig. S1). Shorting the gap in the inner conductor through the outer conductor precludes the formation of the appropriate mirror currents, eliminating all resonances across a broad spectrum (Fig. 1C), which allows the coil to be detuned with PIN diodes during excitation.

The same idea can be leveraged to suppress the current induced by the NMR phenomenon during signal reception.  This suppression of currents represents a fundamental difference between LICs and HICs. LIC elements require a low impedance at the port for current to flow, whereas a low impedance across the port of a HIC suppresses the induced current (Fig. 1D). Consequently, implementing a ``reverse preamplifier decoupling'' scheme, which creates a low impedance at the port, suppresses all currents in a HIC arrangement, provided that the impedance at the port is significantly lower than the intrinsic impedance of the HIC itself ($\approx 2k\Omega$). Whereas creating a sufficiently low impedance at the LIC port is possible only at one isolated frequency and requires precise fine tuning of the preamplifier interface, reversed preamplifier decoupling in a HIC arrangement suppresses all currents over a wide range of conditions.

A 3D printer (Fortus 360mc, Stratasys, Minnesota) was used to explore the design space of possible coaxial substrate configurations (Fig. S2). The lack of lumped elements \cite{Corea2016,Vasabawala2016}  in combination with the coaxial design \cite{Stengard1997} suggests that a high degree of mechanical flexibility may be possible with HIC coils. Given the possibility to make a more flexible coil in addition to leveraging the desirable decoupling properties of HICs, we settled on a 0.7 mm separation between the inner and outer conductor, the effective resolution limit of our printer for this type of structure, leading to a coil diameter of 8.0 cm when tuned to 123 MHz, the proton Larmor frequency of our 3 Tesla MRI scanner (Siemens, Erlangen, Germany).

To evaluate how well HICs cloak themselves from one another, we placed three identical HIC elements side by side on top of a large conductive phantom (Fig. 2A). MR images were obtained with each coil individually (Fig. 2B), and also with all coils active simultaneously (Fig. 2C). For comparison, three conventional 8-cm-diameter LICs were constructed (Fig. 2D) and used for imaging under identical circumstances. When all coils are active at the same time, the signals measured using LICs are distorted due to coupling (Fig. 2 E vs F), whereas the receive profiles of the HICs remain unaltered (Fig. 2 B vs C).

\begin{figure}[!]
\includegraphics[width=8.6cm ]{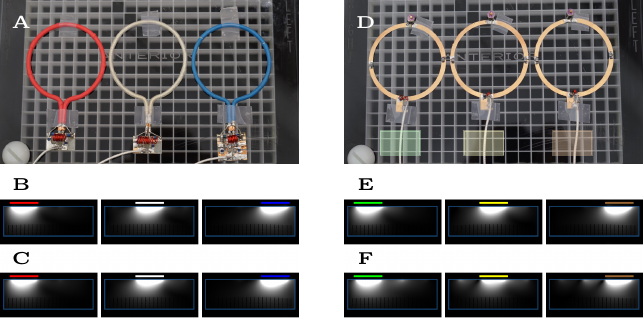}
\caption{Evaluation of signal coupling between three coil elements placed side by side.  Three 8 cm diameter HIC elements (A). MR images obtained with each HIC individually in three separate measurements (B) and with all elements simultaneously active in one single measurement (C).  Three 8 cm diameter LIC elements (D). MR images obtained with each LIC individually in three separate measurements (E) and with all LICs simultaneously active in one single measurement (F). All MR images are displayed on the same scale.}
\end{figure}

Although distinct coil-profiles are a key feature of the NMR phased array, crucial for parallel imaging\cite{Sodickson1997,Pruessmann1999,Griswold2002} and modern multi-band techniques\cite{Larkman2001, Setsompop2012}, decoupling should not come at the expense of a reduced signal to noise ratio (SNR). To investigate this aspect, we evaluated the SNR at 4.2 cm ($\approx$1/2 the coil diameter) depth below the center element as a function of coil overlap (Fig. 3). LIC elements reach their optimal performance when critically overlapped ($\approx$25\% overlap).  When the outer coil elements are moved further apart or closer together, the SNR quickly deteriorates. The HIC elements, on the other hand, show almost no degradation in SNR as the overlap between them is changed (Fig. 3, blue curve).

\begin{figure}[!]
\includegraphics[width=8.6cm ]{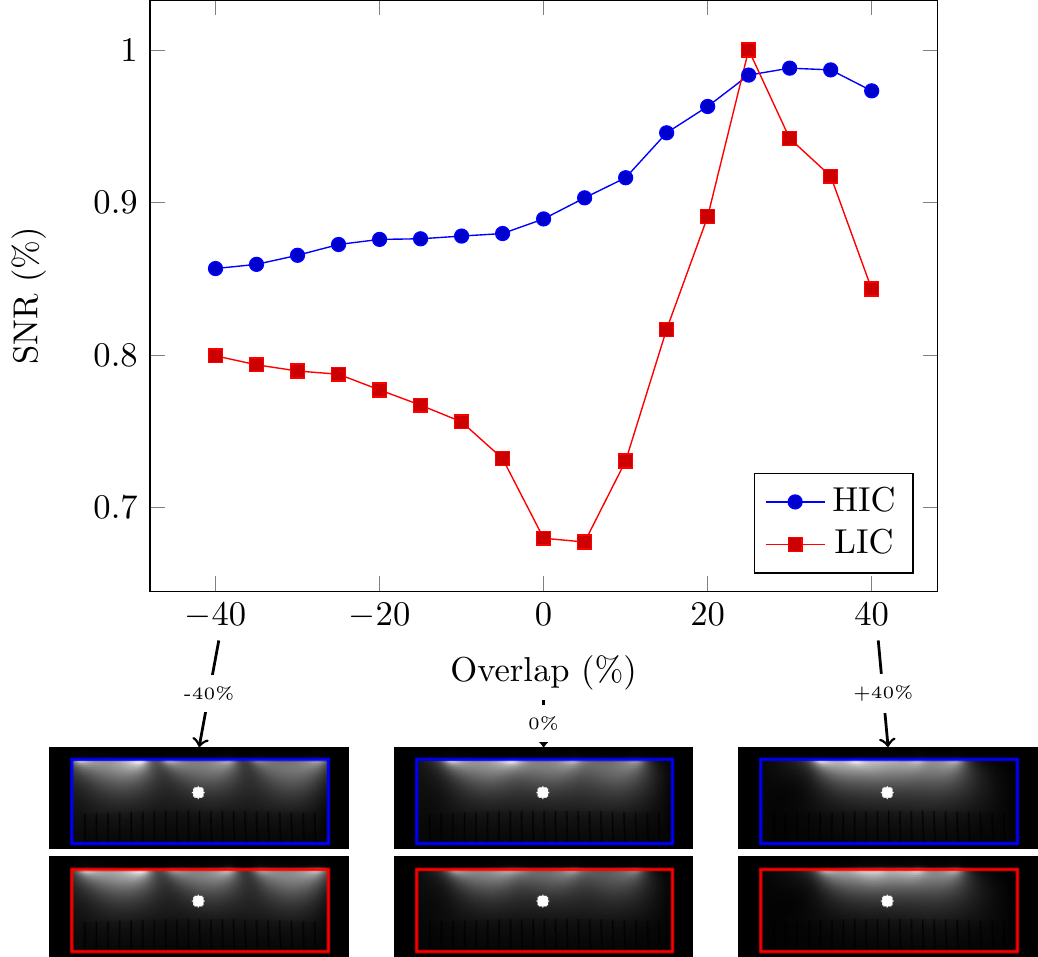}
\caption{Evaluation of SNR degradation due to coupling between neighboring coil elements as a function of coil overlap. The blue (HIC measurements) and red (LIC measurements) lines shows the SNR at 4.2cm below the center coil in a three-coil arrangement. The overlap between coils is varied between -40 and +40 \%. The MR images below the graph show the combined signal images at -40, 0 and +40\% overlap. The blue (HIC data) and red (LIC data) boxes displayed on top of the MR images mark the edges of the phantom. The white dot indicates the position of the SNR measurement.}
\end{figure}

To illustrate the unique degrees of freedom provided by HIC elements, we created a wearable coil in the form of a glove (Fig. 4). In this proof of concept, eight HIC elements were stitched onto a cotton glove. The contours of each finger are traced by an individual coil, allowing all the joints in each finger to articulate freely. Two additional elements were stitched on the top of the hand and wrist, and one additional element was stitched on the bottom, allowing the carpal bones in the wrist joint to be studied as well. 

Figure 4 A shows a coronal and sagittal slice through the hand, stretched out flat on top of the patient table. The close-fitting elements enable fast imaging with exquisite detail (250$\mu$m, 2mm slice). Although the coil elements on each finger are directly adjacent to one another in this configuration, distinct coil profiles are maintained (Fig. 4 C and Fig. 3S). One other unique feature of the glove array, where each individual element traces a distinct anatomical feature, is the ability to isolate individual fingers, simply by selecting the appropriate coil. Alternatively, a factor of 5 in acceleration can be achieved without g-factor penalty\cite{ Pruessmann1999}, by allowing all fingers to alias on top of one another, and relying purely on the coil signal isolation to directly reconstruct the individual images belonging to each finger (Fig. 4S).

The flexibility of the HIC elements, combined with their immunity from electrodynamic coupling, enables the visualization of the intricate dynamics between ligaments, tendons and muscles during complex motions such playing piano/typing (Video S1), or grasping objects (Video S2). Thus, HIC elements open up new avenues for the study of complex joint motion, promising to facilitate, for example, the diagnosis, monitoring and treatment of repetitive strain injuries in athletes, musicians, and others.

Figure 4 B shows the same hand in a different position, now holding a peach, to illustrate how the HIC elements conform to the shape of the hand. Because of the robustness to variations in coil overlap, the performance of the close-fitting coils remains unperturbed, allowing the study of intricate structures such as the pulley and flexor arrangement in different positions and even under load (Fig. S5). Although pulley ligaments and flexor tendon are almost indistinguishable based on underlying contrast (both have short T2 relaxation times), their interaction can now be examined by squeezing an object during the scan (Fig. S5), revealing how the pulley arrangement guides the flexor tendons and distributes the load. Additional high-resolution images (Fig. S6-7) can be found in the supplementary material.

\begin{figure}[h!]
\includegraphics[width=8.6cm ]{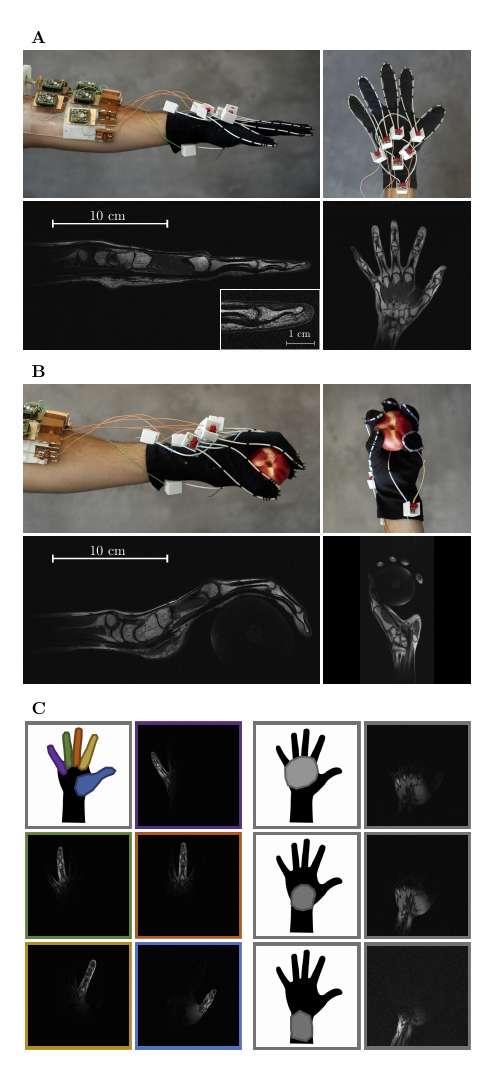}
\caption{Photos of the glove coil with the hand starched out and the corresponding T1 weighted MR images (A). Photos of the glove coil while holding a peach and the corresponding T1 weighted MR images (B). Illustrations showing where the individual coil elements are located on the hand next to the individual coil images obtained during a single T1 weighted measurement with all coils simultaneously activated (C). For a side by side comparison of simultaneously and sequentially acquired images see supplementary figure S3.}
\end{figure}

The proposed HIC elements also have a unique strength when deactivated. Multinuclear MR systems focusing on the less-abundant MR-visible nuclei rely on proton-based MR images to visualize the underlying anatomical structure, which requires the nesting of multiple RF coils operating at different frequencies\cite{Schnall1985,Avdievich2007,Brown2013,Shajan2016}. Whereas traditional LICs can only be detuned over a narrow bandwidth, leading to residual interactions that distort the RF fields and reduce the SNR, HIC elements detune across a wide range of frequencies (Fig. 1C).

The proposed HIC elements pave the way towards size-adjustable and flexible MR coils that allow greater patient comfort and facilitate new areas of research. For example, in addition to enabling the detailed non-invasive study of joint motion \textit{invivo} and pushing the frontiers of multinuclear MR, HIC-based wearable MR coils can boost the study of brain development, by providing comfortable close-fitting high-density coil arrays that adapt to the subject size. In general, by freeing MR from the electromechanical constraints imposed by traditional low-impedance structures, HICs have the potential to bring MR to new areas with a premium on adaptability and a simultaneous need for the highest detector performance.


\subsubsection*{ACKNOWLEDGMENTS}
This work was performed under the rubric of the Center for Advanced Imaging Innovation and Research (CAI2R, www.cai2r.net), a NIBIB Biomedical Technology Resource Center (NIH P41 EB017183). We thank Ryan Bown for critically reading the manuscript, Markus Vester for the many valuable discussions, and Zidan Yu for her help during the experiments. B.Z. build the coils and interfaces. B.Z. and M.A.C design the experiments and collected the data.  B.Z., D.K.S and M.A.C. analyzed the results and wrote the manuscript.

\newpage

\subsection*{Materials and Methods}

\subsubsection*{LIC construction}
The conventional surface coils were constructed as loops 6mm in width and 80mm in diameter, routed out from a 31mil single-sided circuit board (Fig. 1a). Four capacitors were evenly distributed on each loop: two fixed 56pF capacitors (Series 11, Voltronics Corp., Denville, NJ) on either side, a variable capacitor on top for tuning, and an 82pF fixed capacitor for matching at the port. In addition to transforming the impedance of the coil to the optimal noise matching impedance of the preamplifier (Siemens Erlangen, Germany), the 82pF capacitor also forms a detuning trap with a positive-intrinsic-negative (PIN) diode (MA4P4002B-402; Macom, Lowell, MA, USA). The inductor in the trap was hand-wound. The coils were connected to low input impedance preamplifiers (Siemens Medical Solutions, Erlangen, Germany). The length of the cable connecting the coil and the preamplifier was carefully adjusted to provide preamplifier decoupling.

\subsubsection*{HIC construction}
The resonance frequency ($f_0  =  \omega_0 /(2 \pi) $) of the HIC elements was tuned by adjusting the length of the loops ($2\pi r_0$), the relative permittivity of the substrate ($\epsilon_r$), and the ratio between the radius of the outer ($r_1$) and inner radius $r_2$ conductor (Supplemental figure 2). 

The admittance as a function of angular frequency ($Y(\omega)$) can be evaluated numerically by estimating the inductive and capacitive impedance of the HIC element.
When tuned, i.e., open at the port, we may think of the HIC element as two open ended coaxial stubs of length $l$ connected in series by the center conductor. In this case each arm adds a capacitive impedance ($Z_c$) 
\begin{equation}
Z_C(\omega) =- i Z_0 cot(\omega l \sqrt{\epsilon_r}/c)
\end{equation} 
where $Z_0$ the characteristic impedance of the coaxial line. 

For a co-axial cylindrical conductor, the characteristic impendence is given by
\begin{equation}
Z_0(\omega) =  \sqrt{\frac{R+i \omega L_{coax}}{G+i \omega C_{coax}}}
\end{equation}
where $R$ is the resistance per unit length, $G$ is the conductance per unit length of the dielectric, $L_{coax}$ is the inductance per unit length, and $C_{coax}$ is the inductance per unit length. Assuming, $R=0$ and $G=0$, we find $Z_0 = \frac{1}{2 \pi}\sqrt{\frac{\mu_0}{\epsilon_0 \epsilon_r}} ln(\frac{r_1}{r_2})$, where $\mu_0$ is the permeability of free space.

The inductive impedance of the HIC can be approximate by the self inductance of the loop of similar size
\begin{equation}
Z_L(\omega) =  i \omega \mu_0 r_0(ln(\frac{8 r_0}{r_1})-2).
\end{equation} Consequently, the admittance ($Y_{tuned} = 1/Z$) of the HIC can be approximated by\begin{equation}Y(\omega) = (Z_{L }(\omega) + 2 Z_{C}(\omega))^{-1}\end{equation} where the factor 2 accounts for the two open ended stubs in the coil.

To visualize the current distribution on the HIC element, the coil was modeled in CST microwave studio (Darmstadt, Germany). Leveraging the reciprocity relation between transmit and receive fields, the coil was driven using a high impedance voltage source at the port. The pre-amplifier and detuning circuit were excluded from the simulation. The coil was placed in free space surrounded by radiative boundary conditions. The current density distributions were exported and analyzed in Mathematica (Wolfram Research, Champaign, IL).

Computer-aided design of the coil substrate was performed in Solidworks (Dassault Systems, USA).  The substrate was printed on a Fortus 360mc 3D printer (Stratasys, US) using polycarbonate, with a relative permittivity of $\epsilon_r$ = 2.2.  The substrate for each coil was printed in two parts. The inner conductor was made from hand-wound copper threads (13
 strands, final thickness 1mm) and the conductive mesh was taken from a commercial coaxial cable (RG58). First, the inner conductor was placed in the central grove of the substrate. The structure was then capped by adding the second half on top. The coaxial braid was placed around the substrate and stretched out to form a tight coaxial enclosure, taking care that the two halves of the dielectric properly encased the central conductor. Heat shrink tubing around the coaxial structure was used to insulate and protect the main structure. The radius ($r_0$) of the completed HIC loop was 40mm, the radius of the cylindrical inner conductor ($r_2$) was 1mm, the substrate thickness was 0.7mm ($r_1$ = 1.2mm), and the diameter of the coaxial structure before shrink-wrapping was about 2.4mm (Fig. S2). 

The coil was then mounted on a 0.062mil single-sided circuit board (10mmx18mm) made with the same circuit router used to create the LICs described above. The two ends of the outer conductor were connected through the board, and a LC circuit was connected to the two ends of the inner conductor to transform the high impedance of the coil to the optimal noise matching of the preamp (Fig. 1b). The cable length connecting the coil and preamplifier was adjusted to create a low impedance at the port. Note that this is the exact opposite of preamplifier decoupling in traditional low impedance coils, for which a high impedance is arranged at the port.

Interestingly, the above-mentioned impedance transformation can be eliminated when using a high impedance metal-oxide-semiconductor field-effect transistor (MOSFET). Gallium Arsenide MOSFETs, for example, have an optimal noise figure at about 1 to 2 k$\Omega$ (similar to the intrinsic impedance of our HICs). Although removing the transformation network could help to further improve the SNR obtained using the proposed HIC design, we chose to use the same preamplifiers for both LIC and HIC coils to avoid any possible bias due to variations in the preamplifier design.

To detune the coil during transmit, two PIN diodes ware placed between the leads on the inner conductor and outer. When a direct current (DC) is provided, the PIN diodes short the two ends of the inner conductor to the outer conductor (forward bias). During receive, the stray capacitance of the PIN diode is incorporated into the matching network.

In addition to the 3D printed design, a HIC was also implemented using commercially available coaxial cable. RG178 was selected: a 50$\Omega$ cable with dimensions similar to the 3D printed coil design. The RG178 HIC was molded into the same shape as the 3D printed HIC by strapping it to a thin plastic sheet. The same interface, preamplifier and detuning configuration was used.

\subsubsection*{Bench experiments}
A homemade double probe was attached to a network analyzer (Agilent, model E5071C) and the overlap was adjusted to eliminate mutual coupling (less than -70dB). To provide a baseline, the coupling between the two elements in the double probe ($S_{1,2}$) was measured in free space far from any object, over a frequency range of 20MHz to 200MHz. Subsequently, a LIC element was placed underneath the double probe, and the $S_{1,2}$ was measured in the tuned and detuned states. The same process was repeated using a HIC element.

To characterize how the impedance at the port influences the traditional current suppression mechanism of preamplifier decoupling for LICs and the proposed reversed preamplifier decoupling for HICs, a variable resistor was placed across the port of the coil, and the $S_{1,2}$ was measured using the double probe setup. The $S_{1,2}$ was recorded at the Proton Larmor frequency of interest (123Mhz), and was plotted as a function of impedance at the port.

\subsubsection*{Phantom experiments}
To evaluate the interaction between coils and its effect on the SNR, a large cuboid phantom (317x317x104mm) was constructed. The phantom was filled with distilled water, doped with 2.5g/L Sodium chloride (NaCl, Sigma-Aldrich, St Louis, MO, USA) and 50 mg/L manganese (II) chloride tetrahydrate (Cl2Mn 4H20, Sigma-Aldrich, St Louis, MO, USA). The conductivity of the liquid was measured using an Agilent dielectric probe (model 85070E) attached to a network analyzer (model E5071C) calibrated using an eCal module (model 85093C). The conductivity was found to be 0.53 S/m.

Before the SNR experiments, the losses in the coil interface and receive chain were evaluated. A single LIC reference coil was placed on the center of the phantom and connected to the first port on the coil interface. Two gradient recalled echo (GRE) images, axially through the center coil, were acquired, one signal image (flip angle = 25 degrees) and one noise image (flip angle = 0 degrees). Sequence parameters were as follows: 3ms echo time, 200ms repetition time, 256x256 matrix, 384x384mm field of view, 5mm slice thickness, and 300 Hz/pixel readout bandwidth. These measurements were performed for each port on the interface (without moving the coil or interface). The relative signal to noise ratio between receive channels was calculated and incorporated into the subsequent SNR analysis.

Three series of experiments were performed. The first series of experiments was designed to show the effect of signal coupling between elements during receive. Three identical LICs were placed side by side on top of the phantom ($\approx$5mm apart). Four GRE images were acquired: one using all three loops simultaneously, and one for each coil element individually with the other coil elements detuned (same sequence parameters described above). The same experiment was repeated using 3 HICs placed side by side ($\approx$5mm apart). The same axial slice through the center of all three loops was used throughout the experiment.

The second series of experiments was designed to quantify the effect of coil overlap on the SNR. Again, three LIC elements were placed side by side, and this time the overlap between coil elements was varied between -40\% and +40\%, in increments of 5\%, without moving the center element. For each overlap, two gradient echo images were acquired using all three loops simultaneously (same sequence parameters as before), once using a flip angle of 25 degrees, to obtain the signal, and once using a 0 degree flip angle, to collect a noise measurement. In addition, a $B_1^+$ map was acquired using the pre-saturation turbo flash method[30]. The same series of experiments was repeated using three identical HICs. The same axial slice through the center of all three loops was used throughout all measurements. SNR maps were calculated in Matlab as outlined by Kellman \& McVeigh [31]. Receive profiles were derived using a body coil reference image (same sequence parameters), and the optimal SNR coil combination was used to combine the individual images[4]. The $B_1^+$ maps were used to normalize the SNR and remove transmit field variations due to dielectric effects inside the phantom.

Finally, we also measured the SNR obtained with a single LIC, a single 3D printed HIC, and a single RG178 HIC element (same sequence parameters). 

All experiments were performed on a 3 Tesla MR system (Skyra, Siemens, Erlangen, Germany), using the body coil for transmission.

\subsubsection*{Glove coil construction}
Extra long black cotton gloves were used to construct the glove coil. Although our 3D printed coils are flexible, it is not clear how the 3D printed polycarbonate material will withstand repeated bending.  Instead, we elected to use RG178 coaxial cable, which has dimensions similar to our 3D printed design. The RG178 contains a durable Teflon substrate seamlessly encasing the inner conductor. Eight elements were constructed, each with a 292mm circumference. Heat shrink was used to insulate the break in the conductive braid on the far end of the coil. The same interface boards as described above were used to mount the detuning circuit and the impedance transformation circuit. Five coils were stitched along the contours of the fingers, one for each finger. Two additional coils were stitched on the top of the glove, one centered on the hand, and the second centered on the wrist. The final coil was stitched on the bottom of the glove, partially covering the wrist and hand.

The preamplifiers (Siemens, Erlangen, Germany) were mounted on a bracelet fashioned out of acrylic plastic and connected to the MR system using a vendor specific ODU connector (ODU-USA, Camarillo, CA, USA). Two cable traps were placed on the cable connecting the preamplifiers to the scanner. The glove coils were connected to the preamplifiers through a shield cable trap using the same thin coaxial cables found in the cable provided by ODU.

Before in vivo experiments, the detuning circuits were tested on the bench and using phantom experiments in the scanner. During the phantom experiments the difference in transmit reference voltage, with and without the glove coil present, was found to be less than 1\%, and no noticeable distortions could be observed in the transmit field distribution. An infrared camera (E60bx, FLIR Systems, Wilsonville, USA) was used to image the surface temperature of the glove and interface before and after a 15min interval of high applied power (100\% specific absorption rate).  No significant temperature differences were observed ($\Delta T\ll5C^\circ$).

\subsubsection*{\textit{In vivo} experiments}
High resolution T1 weighted images of the left hand were acquired at 3 Tesla using the body coil for transmission and the glove coil for reception. The volunteer was imaged in a prone position on the scanner bed, with the left hand stretched out in front of her. In the first experiment, the hand was placed flat on the table. Low resolution GRE images were used to localize the hand inside the scanner. High resolution (250$\mu$m) T1 weighted turbo spin echo (TSE) images of the left hand were acquired in coronal and sagittal planes (TSE, 1024x786 matrix, 256x192mm field of view, 2mm slice thickness, Turbo factor 2, excitation/refocusing angle of 90/180 degrees, TR =  400 ms, TE = 15ms, total scan time 2min:37sec). In addition, 150$\mu$m resolution images were acquired in sagittal only (TSE, 2048x512 matrix, 303x76mm field of view, 2mm slice thickness, Turbo factor 2, excitation/refocusing angle of 90/180 degrees, TR =  400 ms, TE = 15ms, total scan time 1min:42sec) In the second experiment, with the same general setup, the volunteer was holding a peach. After localization, a second series of coronal and sagittal T1 weighted TSE images was acquired (same sequence parameters as before). In addition, a proton density weighted 3D GRE dataset was acquired covering only a single finger (0.5 mm isotropic, 512x64x104 matrix, 256x32x52 mm field of view, TR = 12ms, TE = 5ms, 10 degree flip angle, total scan time 1min 20sec).

A second volunteer was scanned at 3 Tesla to capture the dynamics of a hand moving inside the scanner. The volunteer was imaged in a prone position on the scanner bed, with the left arm in front. During the first experiment, the volunteer was asked to move her fingers as if playing piano or typing. While the subject moved, 4000 radial projections were acquired in a coronal plane through the hand (Golden angle radial GRE, 0.8x0.8mm, 192x192 matrix, 160x160mm field of view, 2.5mm slice thickness, TR = 6.5ms, TE =3.5ms, total acquisition time 25 seconds). From these 4000 projections, 260 images were reconstructed at 10 frames per second, using a sliding window reconstruction spanning 100 projections[32].

In a second experiment, a peach was placed within reach of the volunteer, on top of the patient table. An axial slice was positioned through the center of the peach. Because there are no coil elements on the peach, the laser positioning system on the scanner was used to locate the appropriate slice. During dynamic MR data acquisition (using the same golden angle radial sequence as before), the subject reached out to grasp the peach. 

The study was approved by our institutional review board (IRB), and written informed consent was obtained prior to the examination.

\onecolumn
 \newpage  
 \begin{figure*}[h!]
\begin{center}
\includegraphics[width=16cm ]{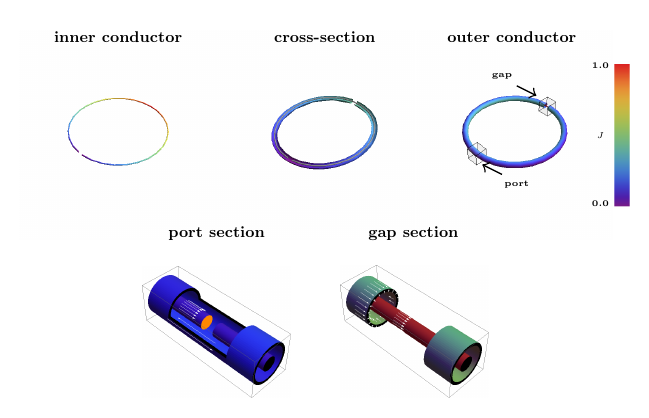}
\end{center}

{ \textbf{Supplemental Figure 1:} The current density (\textbf{J}) distribution on a simplified 8cm diameter HIC coil. Simulations were performed using CST microwave studio (Darmstadt, Germany), assuming radiative boundary conditions and perfect conductors. The top left panel shows the inner conductor only. The top center panel shows a cross-section revealing the current distribution on the inner surface of the outer conductor. The top right panel shows current distribution on the outer surface of the outer conductor. The bottom left shows the gap in the inner conductor at the port. In the simulation the outer conductor completely encases the inner conductor, for displaying purposes only half the outer conductor is shown. The bottom right figure shows the flow of current around the gap in the outer conductor. White arrows indicate the direction of the current flow.}
\end{figure*}

\newpage
 \begin{figure*}[h!]
\begin{center}
\includegraphics[width=8.6cm ]{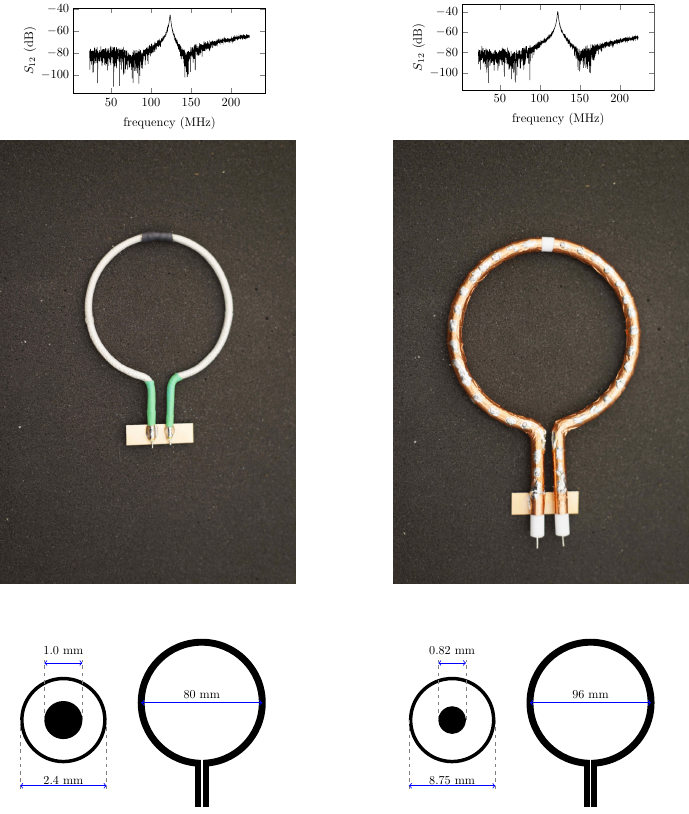}
\end{center}

{ \textbf{Supplemental Figure 2:} Comparison of two 3D printed HIC elements resonating at the same frequency. The ratio between the inner and outer diameter influences the length needed to achieve the resonance frequency. The top row shows the $S_{1,2}$ trace measured on the bench using a double-probe.}
\end{figure*}

\newpage
 \begin{figure*}[h!]
\begin{center}
\includegraphics[width=8.6cm ]{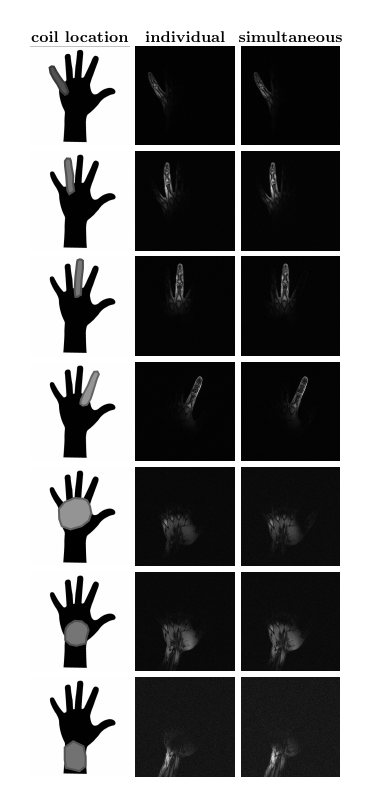}
\end{center}

{ \textbf{Supplemental Figure 3:} Evaluation of signal coupling between HIC elements in the glove coil. Gradient recalled echo images acquired in 8 separate measurements, each with only one coil activated at a time (center column). Gradient recalled echo images acquired in one single measurement with all coils activated at the same time (right column). Each row is displayed at the same scale. Sequence parameters: TR=10ms, TE=3.72ms. flip angle = 25, 512x512 Matrix, 265x265mm FOV, 2mm slice thickness, 0.5x0.5mm resolution, total scan time = 5 s.}
\end{figure*}

\newpage
 \begin{figure*}[h!]
\begin{center}
\includegraphics[width=12.0cm ]{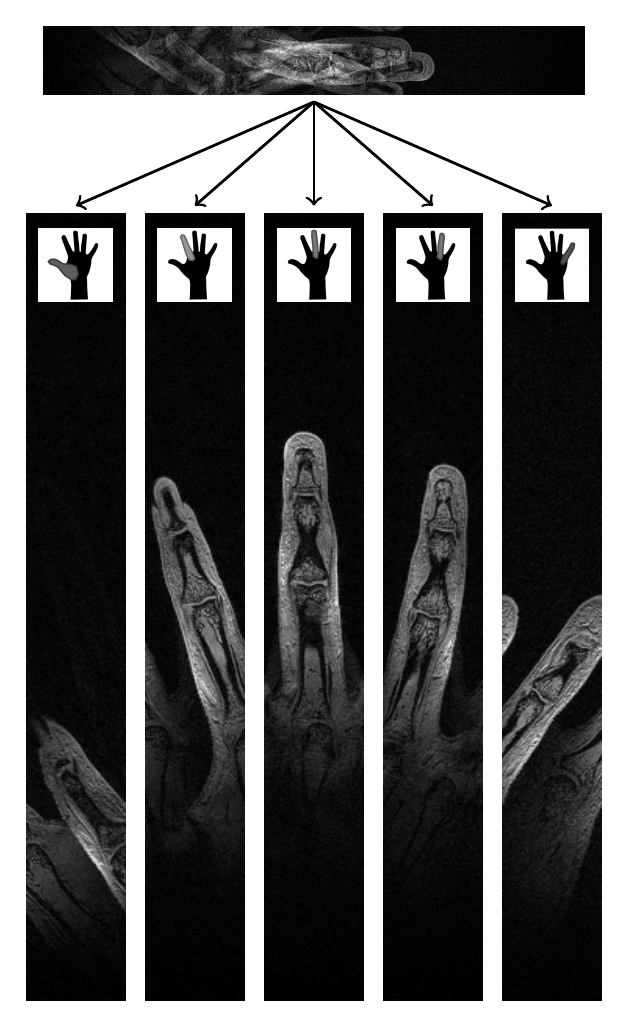}
\end{center}

{ \textbf{Supplemental Figure 4:} A 3D data set (non-selective excitation) acquired covering only the middle finger. The top panel shows a central slice from the 3D data set combining all data from all coils. Underneath the top panel, each of the individual coil images (measured simultaneously) are placed side by side to reveal  the individual fingers. Note that, no parallel imaging reconstruction techniques were used to separate the images. (0.5 mm isotropic, 512x64x104 matrix, 256x32x52 mm field of view, TR = 12ms, TE = 5ms, 10 degree flip angle, total scan time 1min 20sec).}
\end{figure*}

\newpage
 \begin{figure*}[h!]
\begin{center}
\includegraphics[width=16.0cm ]{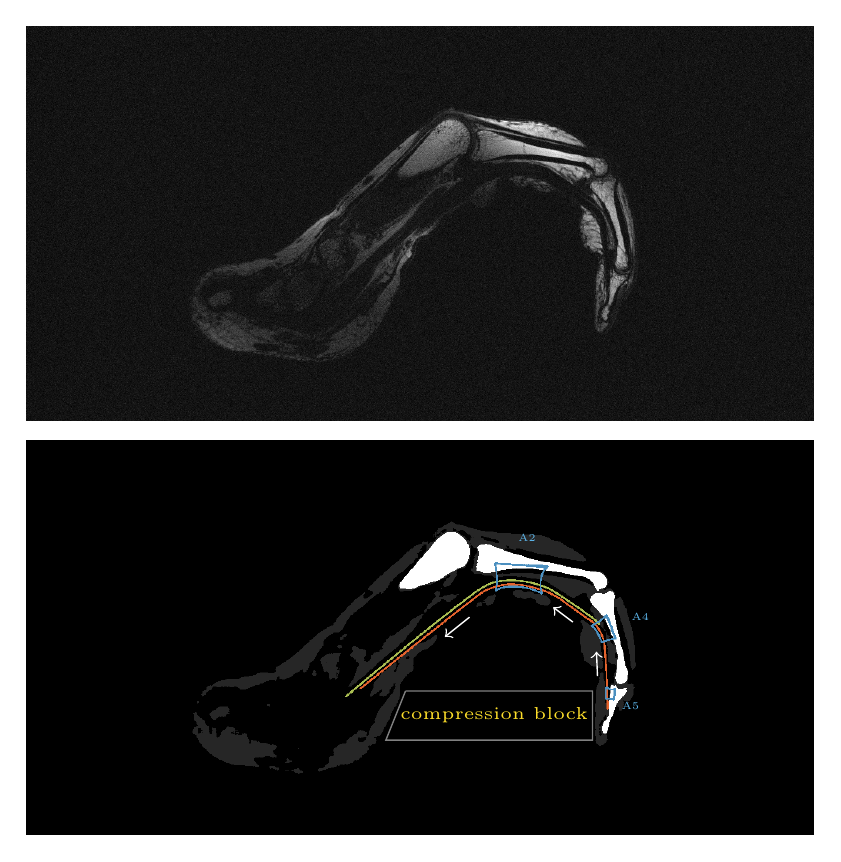}
\end{center}

{ \textbf{Supplemental Figure 5:} Sagittal T1 weighted image through the flexor tendon in the index finger while squeezing a block of wood (top penal). Although pulley ligaments and flexor tendon are almost indistinguishable based on underlying contrast (both have short T2 relaxation times), their interaction can now be revealed. As the subject squeezes on the compression block, the load on the flexor tendon is distributed over the pulleys as illustrated in the bottom panel.}
\end{figure*}

\newpage
 \begin{figure*}[h!]
\begin{center}
\includegraphics[width=17cm ]{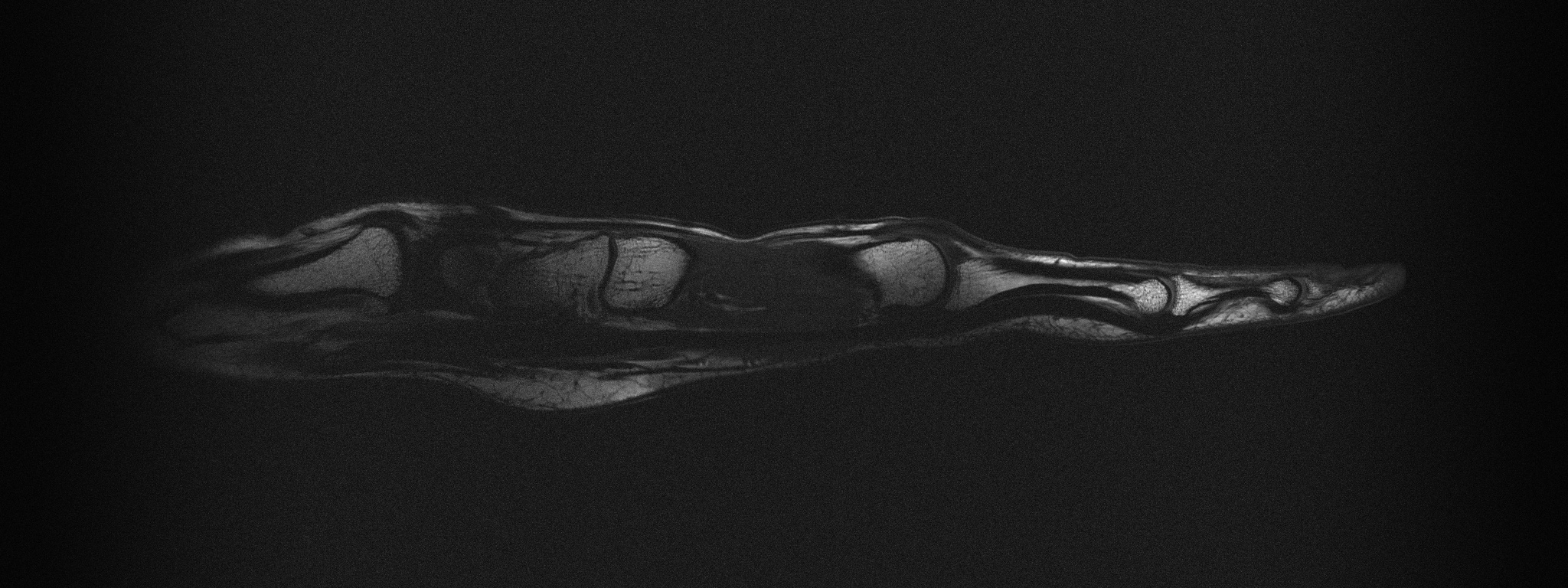}
\end{center}

{ \textbf{Supplemental Figure 6:} Sagittal T1 weighted image through the flexor tendon in the index finger while the hand is stretched out. Sequence parameters: TR=400ms, TE=15ms, excitation angle = 90, refocusing angle = 180, Turbo factor 2, 2024x512 Matrix, 303.6x76.8mm FOV, 2mm slice thickness, 150x150 $\mu$m resolution, total scan time = 1min  42s.}
\end{figure*}

 \begin{figure*}[h!]
\begin{center}
\includegraphics[width=17cm ]{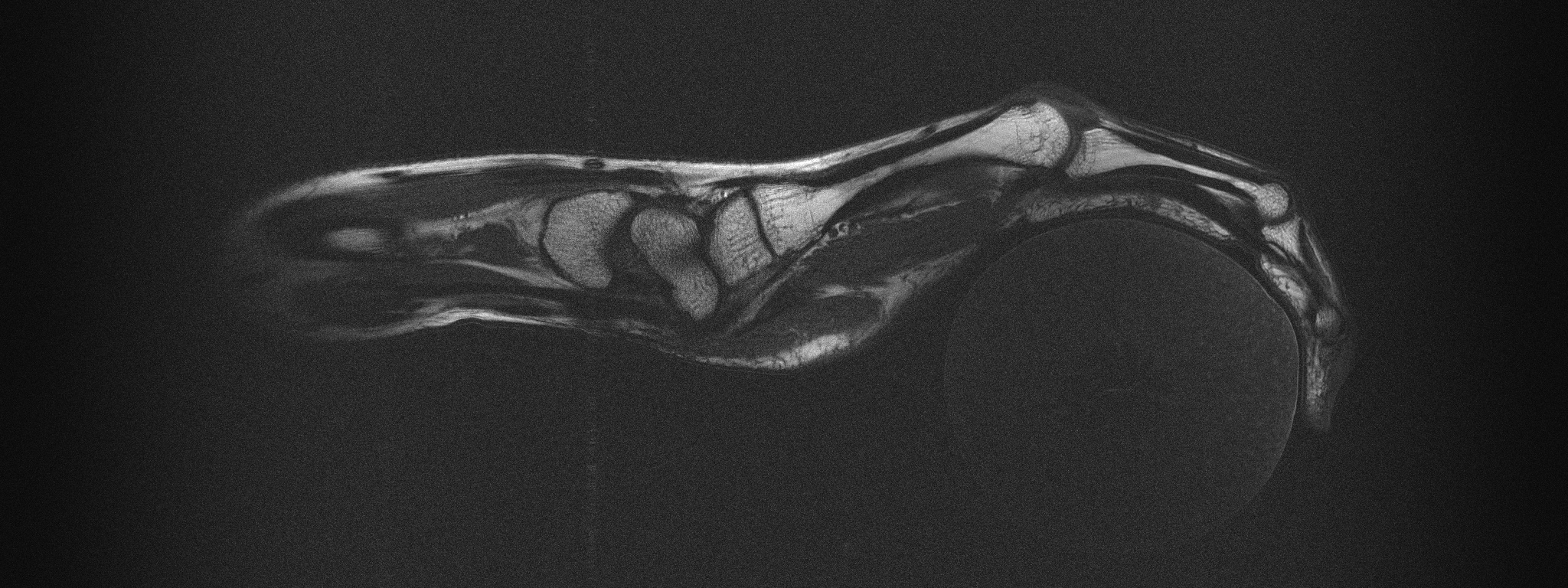}
\end{center}

{ \textbf{Supplemental Figure 7:}  Sagittal T1 weighted image through the flexor tendon in the index finger while the hand is holding a peach. Sequence parameters: TR=400ms, TE=15ms, excitation angle = 90, refocusing angle = 180, Turbo factor 2,  2024x512 Matrix, 303.6x76.8mm FOV, 2mm slice thickness, 150x150 $\mu$m resolution, total scan time = 1min  42s.}
\end{figure*}

\newpage
 \begin{figure*}[h!]
\begin{center}
\includegraphics[width=16.0cm ]{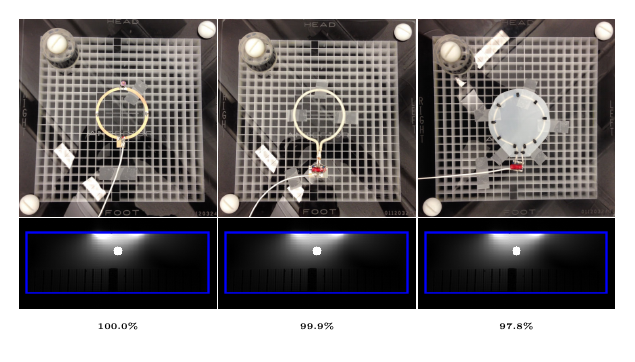}
\end{center}

{ \textbf{Supplemental Figure 8:} Evaluation of the signal to noise ratio relative to a traditional 8cm low impedance coil element. From left to right: 8 cm low impedance coil, 8cm 3D printed high impedance coil, and a high impedance coil fabricated from RG178. The number underneath each column indicates the relative SNR measured at 42mm depth compared to the 8cm traditional low impedance coil.}
\end{figure*}

\newpage
{ \textbf{Supplemental Video 1:} MR video showing a coronal slice through the left hand while the subject moves her fingers in a motion pattern similar to playing the piano or typing on a keyboard (0.8x0.8mm in-plane resolution, 2.5mm slice thickness).}

\vspace{1cm}
{ \textbf{Supplemental Video 2:} MR video of the left hand grabbing an object (peach). The imaging plane is centered on the peach, but there are no MR coils on the peach itself. As the hand wearing the glove coil moves through the imaging plane different parts of the hand become visible. As the hand grabs the object the coils on the glove also wrap the peach, revealing the inner structure of the fruit (0.8x0.8mm in-plane resolution, 2.5mm slice thickness).}


\begin{thebibliography}{9}
\small
\bibitem{Bloch1948} 
Bloch F., Nicodemus D., Staub H.A., Quantitative Determination of the Magnetic Moment of the Neutron in Units of the Proton Moment. Phys. Rev., 1948;74:1025.
\bibitem{Lauterbur1973} 
Lauterbur  P.C., Image Formation by Induced Local Interactions: Examples Employing Nuclear Magnetic Resonance. Nature 1973;242:190-191.
\bibitem{Mansfield1973} 
Mansfield, P., Grannell, P.K., NMR 'diffraction' in solids? Phys. C. Solid. Stat., 1973; L422-L426.
\bibitem{Roemer1990} 
Roemer P.B., et al., The NMR Phased Array. Magn. Reson. Med., 1990;16:192-225.
\bibitem{Wang1995} 
Wang J., Reykowski A., Dickas J., Calculation of the signal-to-noise ratio for simple surface coils and arrays of coils. IEEE Trans. Biomed. Eng. 1995;42:908-917.
\bibitem{Ocali1998} 
Ocali O., Atalar E., Ultimate intrinsic signal-to-noise ratio in MRI. Magn. Reson. Med., 1998;39:462-473.
\bibitem{Schnell2000} 
Schnell W., Renz W., Vester M., Ermert H., Ultimate Signal-to-Noise Ratio of Surface and Body Antennas for Magnetic Resonance Imaging. IEEE Trans. Antennas. Propag. 2000;48:418-428.
\bibitem{Ohliger2003} 
Ohliger M.A., Grant A.K., Sodickson D.K., Ultimate intrinsic signal-to-noise ratio for parallel MRI: Electromagnetic field considerations. Magn. Reson. Med., 2003;50:1018-1030.
\bibitem{Wiesinger2004} 
Wiesinger F., Boesiger P., Pruessmann K.P., Electrodynamics and ultimate SNR in parallel MR imaging. Magn. Reson. Med., 2004;52:376-390.
\bibitem{Lattanzi2010} 
Lattanzi R., et al.,. Performance evaluation of a 32-element head array with respect to the ultimate intrinsic SNR. NMR Biomed 2010;23:142-151.
\bibitem{Lattanzi2012} 
Lattanzi R, Sodickson DK. Ideal current patterns yielding optimal signal-to-noise ratio and specific absorption rate in magnetic resonance imaging: computational methods and physical insights. Magn. Reson. Med., 2012;68:286-304.
\bibitem{Wiggins2009} 
Wiggins G.C., et al., 96-Channel receive-only head coil for 3 Tesla: design optimization and evaluation. Magn. Reson. Med., 2009;62:754-62.
\bibitem{Schmitt2008} 
Schmitt M, et al. A 128-channel receive-only cardiac coil for highly accelerated cardiac MRI at 3 Tesla. Magn. Reson. Med., 2008;59:1431-1439.
\bibitem{Fujita2013} 
Fujita H., Zheng T., Yang X., Finnerty M.J., Handa S., RF surface receive array coils: the art of an LC circuit. J. Magn. Reson. Imaging, 2013;38:12-25.
\bibitem{Kurs2007} 
Kurs A., et al., Wireless Power Transfer via Strongly Coupled Magnetic Resonances. Science, 2007;317:83-86.
\bibitem{Tierney2014} 
Tierney B., Grbic A., Planar shielded-loop resonators for wireless non-radiative power transfer. IEEE Antennas and Propagation Society International Symposium (APSURSI). 2014;DOI:10.1109/APS.2013.6711080.
\bibitem{selfMethods}
{Information on materials and methods in the supplements following the main body of the paper.}
\bibitem{Corea2016} 
Corea J.R, et al., Screen-printed flexible MRI receive coils. Nat. Commun. 2016;7:10839.
\bibitem{Vasabawala2016} 
Vasanawala A., et al, Development and Clinical Implementation of Very Light Weight and Highly Flexible AIR Technology Arrays. Proc. Intl. Soc. Mag. Reson. Med., 2017;0755 
\bibitem{Stengard1997} 
Stengard A., Planar Quadrature Coil Design Using Shielded-Loop Resonators. J. Magn. Reson., 1997;125:84-91.
\bibitem{Sodickson1997} 
Sodickson D.K., Manning  W.J. Simultaneous acquisition of spatial harmonics (SMASH): fast imaging with radiofrequency coil arrays. Magn. Reson. Med., 1997;38:591-603.
\bibitem{Pruessmann1999} 
Pruessmann K., Weiger M, Scheidegger MB, Boesiger P. SENSE: sensitivity encoding for fast MRI. Magn. Reson. Med., 1999;42:952-962.
\bibitem{Griswold2002} 
Griswold M., et al., Generalized autocalibrating partially parallel acquisitions (GRAPPA). Magn. Reson. Med.,  2002;47:1202-1210. 
\bibitem{Larkman2001}
Larkman D.J., Hajnal J.V., Herlihy A.H., Coutts G.A., Young I.R., Ehnholm G. Use of multicoil arrays for separation of signal from multiple slices simultaneously excited. J Magn. Reson. Imaging. 2001;13(2):313-317.
\bibitem{Setsompop2012} 
Setsompop K., et al., Blipped-controlled aliasing in parallel imaging for simultaneous multislice echo planar imaging with reduced g-factor penalty. Magn. Reson. Med., 2012;67:1210-1224.
\bibitem{Schnall1985}
Schnall M.D., Harihara Subramanian V., Leigh J.S., Chance B.. A new double-tuned probed for concurrent 1H and 31P NMR. J. Magn Reson. 1985;65:122-129.
\bibitem{Avdievich2007}
Avdievich NI, Hetherington HP. 4 T Actively detuneable double-tuned 1H/31P head volume coil and four-channel 31P phased array for human brain spectroscopy. J Magn Reson. 2007;186:341-346. 
\bibitem{Brown2013}
Brown R., et. al.,  Design of a nested eight-channel sodium and four-channel proton coil for 7T knee imaging. Magn. Reson. Med., 2013; 70:259-268. 
\bibitem{Shajan2016}
Shajan G., et al.,Three-layered radio frequency coil arrangement for sodium MRI of the human brain at 9.4 Tesla. Magn. Reson. Med., 2016;75:906-916. 
\bibitem{Chung2010}
Chung S., Kim D., Breton E., Axel L. Rapid B1+ mapping using a preconditioning RF pulse with TurboFLASH readout.
 Magn. Reson. Med., 2010;64:439-446.
\bibitem{kellman2005} 
Kellman P, McVeigh E.R. Image reconstruction in SNR units: a general method for SNR measurement. Magn. Reson. Med., 2005;54:1439-1447.
\bibitem{Winkelmann2007} 
Winkelmann S., Schaeffter T., Koehler T., Eggers H., Doessel O. An optimal radial profile order based on the Golden Ratio for time-resolved MRI. IEEE Trans. Med. Imaging. 2007;26:68-76.
\end{thebibliography}
\end{document}